\documentclass{elsart}
\usepackage{graphicx}
\usepackage{epsfig}
\usepackage{amsmath}
\usepackage{amssymb}
\newcommand{\Au} {\mbox{$ ^{197}{\rm{Au}}$}~}

\newcommand{\neut}{$\tilde{\chi}$~}

\newcommand{\Co} {\mbox{$ ^{57}{\rm{Co}}$}~}

\newcommand{\hetrois}    {\mbox{$ ^{3}{\mathrm{He}}                            $}~}
\newcommand{\hetro}    {\mbox{$ ^{3}{\mathrm{He}}                            $}}

\newcommand{\gam}{\mbox{\rm $\gamma$-ray}} 
 
\newcommand{\xrays}{\mbox{\rm X-rays}} 
\newcommand{\gams}{\mbox{\rm $\gamma$-rays}}

\newcommand{\micron}{\mbox{{\rm$\mu$m}}}
\newcommand{\microns}{\mbox{{\rm$\mu$m}}~}
\newcommand{\muK}{\mbox{{\rm $\mu$}K}}

\def\NIMA#1#2#3{{\rm Nucl.~Instr.~Methods} {\bf{A#1}} (#2) #3}

\def\PLB{{\em Phys. Lett.}  B}

\def\PRD#1#2#3{{\rm Phys. Rev.} {\bf{D#1}} (#2) #3}
\def\PRB#1#2#3{{\rm Phys. Rev.} {\bf{B#1}} (#2) #3}

\def\PRL#1#2#3{{\rm Phys.~Rev.~Lett.} {\bf{#1}} (#2) #3}
\def\PR#1#2#3{{\rm Phys.~Rept.} {\bf{#1}} (#2) #3}
\def\PLB#1#2#3{{\rm Phys.~Lett.} {\bf{B#1}} (#2) #3}

\def\APJ#1#2#3{{\rm Astrophys.~J.} {\bf{#1}} (#2) #3}
\def\APJS#1#2#3{{\rm Astrophys.~J.~Suppl.} {\bf{#1}} (#2) #3}
\def\AA#1#2#3{{\rm Astron. \& Astrophys.} {\bf{#1}} (#2) #3}

\def\NATURE#1#2#3{{\rm Nature} {\bf{#1}} (#2) #3}

\usepackage{amssymb}

\begin{document}
\begin{frontmatter}
\title{Low energy conversion electron detection in superfluid \hetrois at ultra-low temperature}
\author{E. Moulin\thanksref{corr1}\thanksref{lpsc}},
\thanks[corr1]{Corresponding author : Emmanuel.Moulin@lpsc.in2p3.fr (phone: +33 4 76 28 41 49, fax:+33 4 76 28 00 04)}
\author{C. Winkelmann\thanksref{crtbt}},
\author{J. F. Mac\'{\i}as-P\'erez\thanksref{lpsc}},
\author{Yu. M. Bunkov\thanksref{crtbt}},
\author{H. Godfrin\thanksref{crtbt}},
\author{D. Santos\thanksref{corr2}\thanksref{lpsc}}
\thanks[corr2]{Corresponding author : Daniel.Santos@lpsc.in2p3.fr (phone: +33 4 76 28 40 21, fax:+33 4 76 28 00 04)}
\address[lpsc]{Laboratoire de Physique Subatomique et de Cosmologie, 
CNRS and Universit\'e Joseph Fourier, 53, avenue des Martyrs, 38026 Grenoble cedex, 
France}
\address[crtbt]{Centre de Recherches sur les Tr\`es Basses Temp\'eratures,
CNRS and Universit\'e Joseph Fourier, BP166, 38042 Grenoble cedex 9, France}

\begin{abstract}
We report on the first results of the MACHe3 (MAtrix of Cells of Helium 3) prototype experiment concerning the measurement of 
low energy conversion electrons at ultra-low temperature. 
For the first time,
the feasibility of the detection of low energy electrons is demonstrated in superfluid \hetro-B 
cooled down to 100 \muK. Low energy electrons at 7.3 keV coming from the K shell conversion of the 14.4 keV 
nuclear transition of a low activity \Co source are detected, opening the possibility 
to use a \hetrois-based detector for the detection of Weakly Interacting Massive Particles
(WIMPs) which are expected to release an amount of energy higher-bounded by 5.6~keV. 
\end{abstract}

\end{frontmatter}
\section{\label{sec:intro}Introduction}
Recent CMB anisotropies measurements \cite{wmap,archeops} used in combination  
with large scale structure surveys~\cite{tegmark} and type Ia supernova measurements \cite{snap} suggest
that roughly 85\% of the matter content of the Universe is
composed of non-baryonic cold dark matter. 
Weakly Interacting Massive Particles (WIMPs) are well justified candidates for this cold dark matter for which  
supersymmetric theories (SUSY) provide a compelling candidate~\cite{jungman}.
Under the Minimal Supersymmetric Standard Model assumption~\cite{jungman}, the Lightest Supersymmetric Particle
(LSP) is found to be the lightest neutralino, the lowest mass neutral 
and colorless linear combination of the superpartners of the gauge and Higgs bosons \cite{jungman}. 
Many promising detectors \cite{edelweiss,cdms} have been developed
to search for non-baryonic cold dark matter. These detectors have reached 
sufficient sensitivity to be able to test regions of the SUSY parameter space, 
but they share common 
problems such as the rejection of the neutron interactions, the radioactive contamination 
and neutrons induced by high-energy muon interactions in the shielding. \\
Following early experimental
works \cite{lancaster,bunkov}, a high granularity detector using superfluid \hetrois 
as sensitive medium has been proposed~\cite{dm2000,nima,4micc}. The use of
\hetrois is motivated by the following properties:
i) As \hetrois is a spin 1/2 nucleus, a \hetrois detector will 
be mainly sensitive to the axial interaction, making this device complementary to 
existing ones, mostly sensitive to the scalar interaction. The axial
interaction is dominant, in general by two orders of magnitude, in the SUSY region associated
with a WIMP-proton elastic cross-section~\cite{plbf}.
ii) A close to absolute purity (nothing can be dissolved in \hetrois at 100 $\mathrm{\mu}$K)~\cite{nature95}.
iii) A high neutron capture cross-section
leading to a well defined signature ${\rm 
n\,+\,\hetrois\,\rightarrow\,p\,+^3H\,+\,764\,keV}$, easy to be detected.
Neutron interaction has thus a clear signature~\cite{dm2000,nima,nature95} and
is well discriminated from a WIMP signal.  
iv) Low Compton cross-section, two orders of magnitude weaker than Ge.
v) No intrinsic X-rays.
vi) A high integrated signal to noise ratio, due to the narrow energy range 
expected for a WIMP signal.
For massive WIMPs the maximum recoil energy depends very weakly on the
WIMP mass.
As the \hetrois target nucleus (${\rm m=2.81 \, 
GeV}\!/c^2$) is much lighter than the WIMP mass
the energy range is bounded by ${\rm 5.6 \,keV}$~\cite{dm2000,nima,plbf}.
To confirm that events in this low energy range can be detected,
an electron consersion source of \Co has been spot-welded in one of the cells of 
a three cell prototype, MACHe3 (MAtrix of Cells of Helium 3). 
We report here the results from the prototype concerning the detection of a few keV electrons
at ultra-low temperature ($\sim$100 \muK) in superfluid \hetro-B. \\
\section{\label{sec:prototype}The MACHe3 prototype}
The elementary component of the MACHe3 prototype is a bolometer cell consisting of a cylindrical copper box, 
with 500 \microns thick walls, 
filled with superfluid \hetro. The box is in weak thermal contact with the outer bath 
through a 200 \microns diameter orifice.
As a particle interacts inside the cell, the released energy is converted into quasiparticles.
The time constant for thermal relaxation 
of quasiparticles through the orifice is controlled by the geometry of the box and the 
size of the orifice. This is tuned to be of the order of 5~s. 
Each cell contains at least one vibrating wire resonator (VWR) 
consisting of a 4.5 \microns diameter superconducting NbTi wire. This forms a semi-loop oscillating 
perpendicularly to its plane at 500 Hz frequency in a 100 mT magnetic field. 
The frequency of the vibration is measured via the 
voltage induced by the motion of the wire through the field lines.
The friction with the quasiparticle cloud induces a damping in the resonance
which is  related to the energy deposited by the incoming particle. \\
A working prototype has been developed using three elementary cells,
one of which contains a very low activity \Co source ($\simeq$ 0.06 Bq).
With this prototype, the muon background discrimination using the correlation among 
the cells and  the detection of low energy events have been performed. 
Fig.~\ref{fig:prototype} presents a detailed view of the prototype.
The middle cell has common  25 \microns thick walls  with the adjacent ones. 
Each of them consists of a cylindrical copper box as described below.
The bottom cell contains an extra VWR 
The experimental set-up is immersed in the \hetrois bath at $\sim$ 100 \muK. 
A key feature of the experiment is 
the low activity \Co naked source which has been
deposited on a  25 \microns thick gold foil spot-welded to the inside middle cell wall. 
We obtain the spectrum at low energy from the \Co naked source shown on fig.~\ref{fig:specd5}.
The background to the detection of the electron lines comes from cosmic muons passing across the cells on peripheral tracks
and the ${\rm \gamma}$-ray lines of 14.4, 122 and 136 keV whose intensities are 9.15\%, 85.52\% 
and 10.71\% respectively. 
In order to estimate the background structure we have simulated the expected background contribution  
to the low energy electron spectrum with the Geant4 package~\cite{geant4}. 
The simulation set-up reproduces the main features of the prototype:
the three bolometric cells including the copper walls and the \Co source.
We generate isotropically the \gams~produced 
by the source. Cosmic muons of 2 GeV are generated from their measured flux at sea level~\cite{grieder} following 
a ${\rm \cos^2(\theta)}$ distribution with respect to the azimuth.
In the 1 to 40 keV energy range, 
97\% of the total number of background events are coming from cosmic muons, whereas only 
3\% are attributed to 14.4, 122 and 136 keV $\gamma$ interactions in \hetrois or in the copper walls.
\section{\label{sec:data}Data analysis and simulations}
\indent A data analysis procedure to extract the amplitude of the events presented in the data has been developed. 
It performs an adequate treatment of very low energy events.
The method is based on four steps. First, wavelet denoising allows us to
reduce significantly the noise on the raw data. Second, the baseline, corresponding to
low frequency temperature fluctuations in the thermal bath,
is then removed. Third, for each data set, an isolated peak is selected and extracted from the data. The shape of this peak 
is related to the temperature of the \hetrois bath~\cite{nature95}. As the shape is fixed for all the peaks independently of their
amplitude, the selected peak is used as a reference.  
Finally, to measure the particle deposited energy, an iterative
fit including peak flagging is applied to the data. 
In the first iteration, we flag the high amplitude peaks by searching spikes on the first derivative of the data.
Then, those are fitted using the reference peak and a first estimate of their positions and amplitudes is obtained. 
For the second iteration, the
fit to the data is subtracted from the data and the flagging procedure is repeated so that smaller
amplitude peaks are detected. The new flagged peaks and the previously flagged ones are fitted as above. 
This procedure is repeated until the best possible fit is found by minimizing the $\chi^2$ value.
The final output of this analysis is the position, amplitude and signal to noise ratio, defined as the amplitude
of the peak divided by the rms of the residuals on the fit,
for each detected peak.  
In particular, very low amplitude peaks up to 1 keV can be retrieved once the higher amplitude ones have been
removed. \\
\indent In order to estimate the performance of the above procedure, a simulation of a timeline reproducing the main features 
of the data has been performed. This includes cosmic muons for which the incident flux expected in our prototype
and the energy spectrum can be well estimated from 
experimental data \cite{collin}.
The electron lines coming from the \Co source described before 
have been implemented in the simulation with their corresponding relative intensities 
leading to an accurate description of the low energy contributions in the data. 
The input spectrum has been convolved with a 1 keV FWHM Gaussian corresponding to the
expected resolution of the VWR measurements.
We have added noise to the simulated data either using a
white noise realization with 1 keV standard deviation  or using
an estimate of the noise extracted directly from the data. 
An efficiency of the order of 
90\%~in the 1 to 40 keV energy range is reached concerning the detection of peaks
in the raw data. A selection 
criteria based on the signal to noise ratio has been used to reject badly fitted peaks and
noise. When selecting peaks with signal to noise ratio higher than 1, the 7.3 and 13.6 keV electron lines 
are well separated in the recovered \Co source spectrum. However the spectrum remains dominated by noise for
energies lower than 1 keV. For the case of signal to noise ratios
higher than 5 the noise contribution around 1 keV fully disappears and therefore we can consider that
for higher energies the noise level is negligible. Thus, for the results presented below we impose
a lower cut of 5 in the signal noise ratio of the fitted peaks in order to keep the best fitted ones and reduce 
the noise contribution in the low energy range.
\section{\label{sec:results}Results and discussions}
\begin{figure}[!ht]
\begin{center}
\includegraphics[scale=0.5]{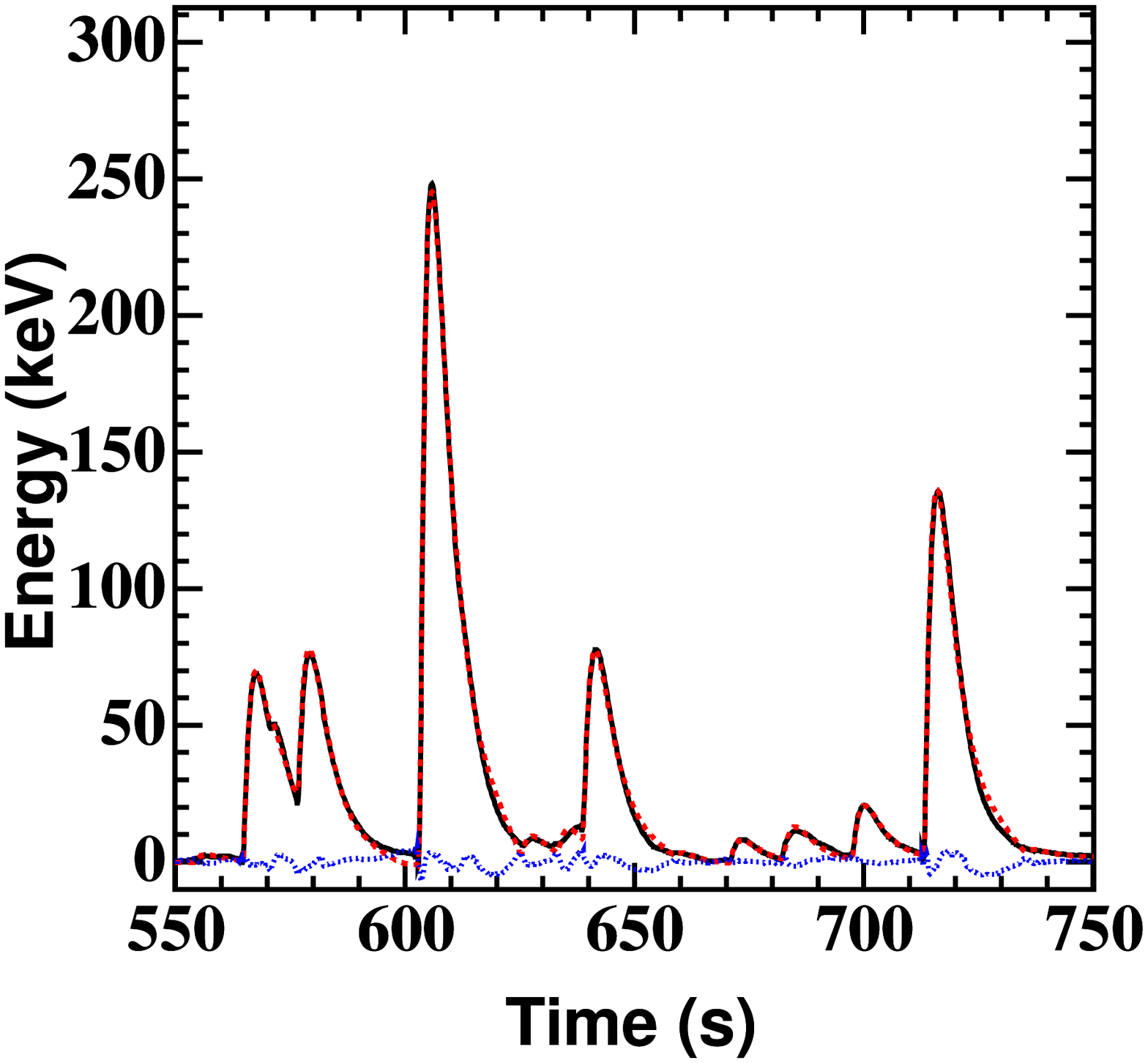}\\
\includegraphics[scale=0.5]{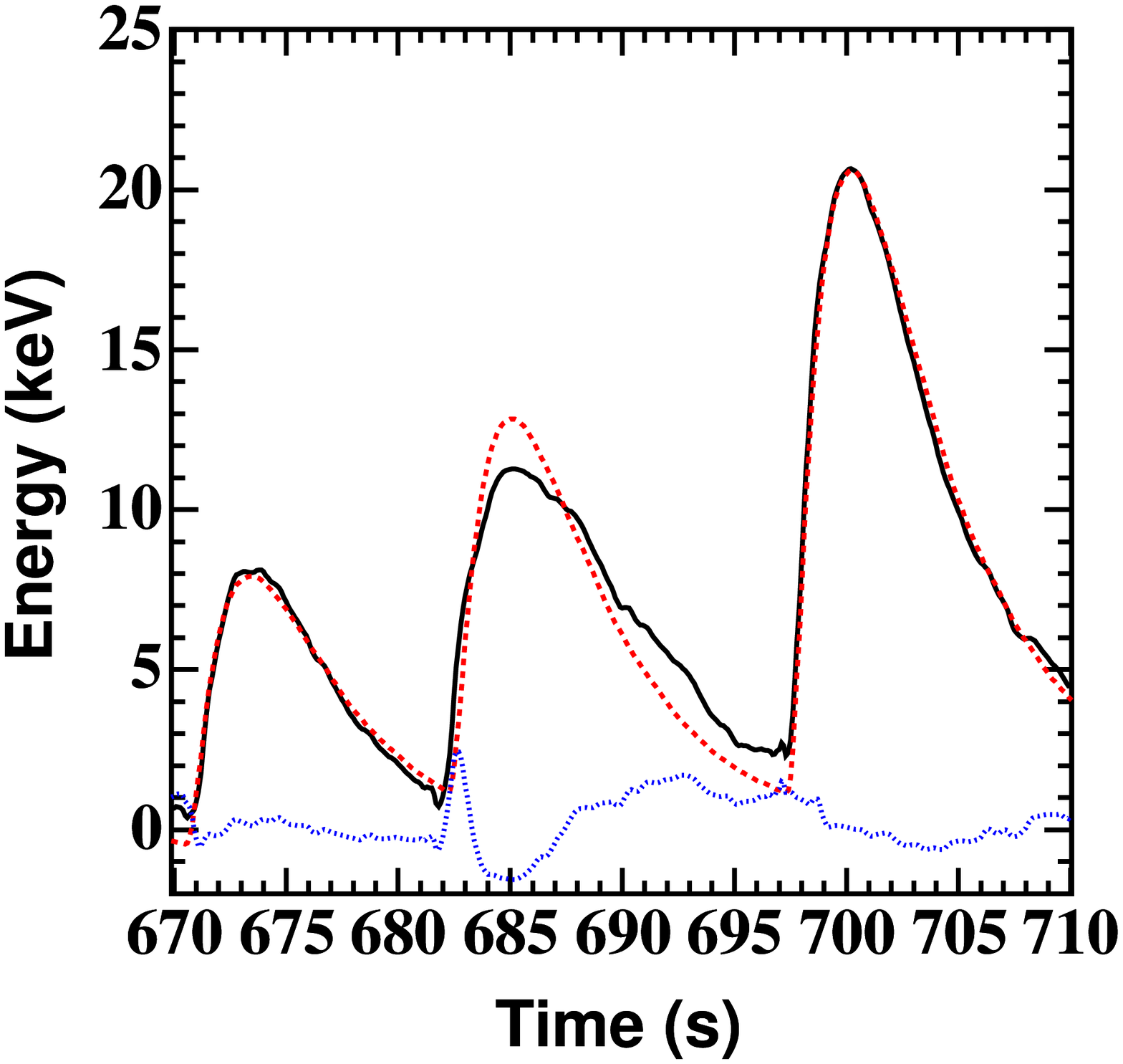}
\caption{\label{fig:rawdata}Energy calibrated frequency width measured by the VWR as a function of time after 
wavelet denoising.
The temperature of the \hetrois bath is close to 100 \muK. 
The observed peaks on the data result from cosmic muons crossing the cell and electrons coming from
the low activity \Co source. Microvibration level is found to be very low.
Solid curve represents denoised data, dashed curve is the fit to data described on the text and the dotted one corresponds to
the residuals for this fit.
We zoom up on low energy events (bottom plot). Electrons coming from the \Co source can be clearly seen indicating that the  
background level is of the order of 1 keV.}
\end{center}
\end{figure}

\begin{figure}[!ht]
\begin{center}
\includegraphics[scale=0.5]{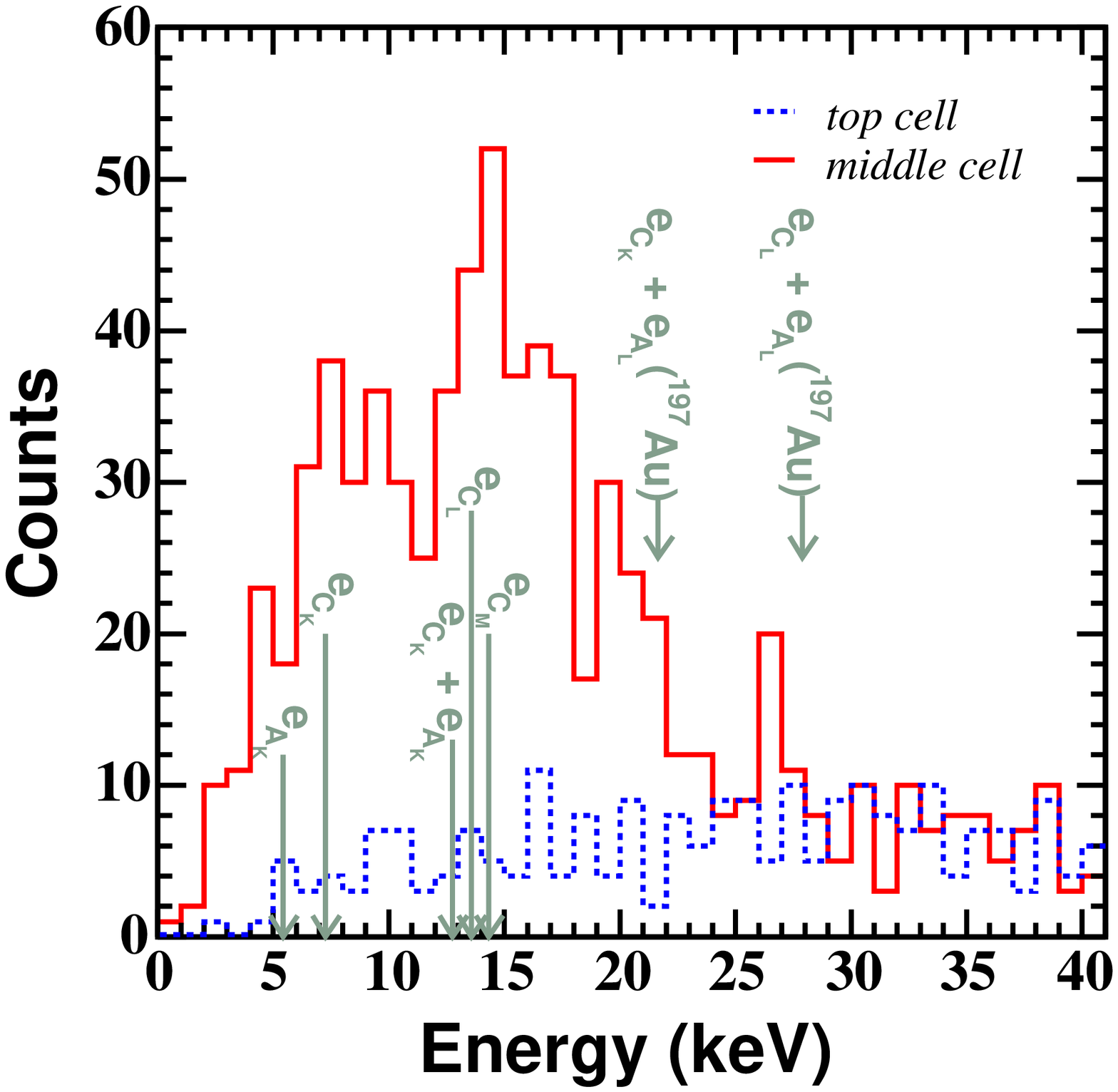}
\includegraphics[scale=0.5]{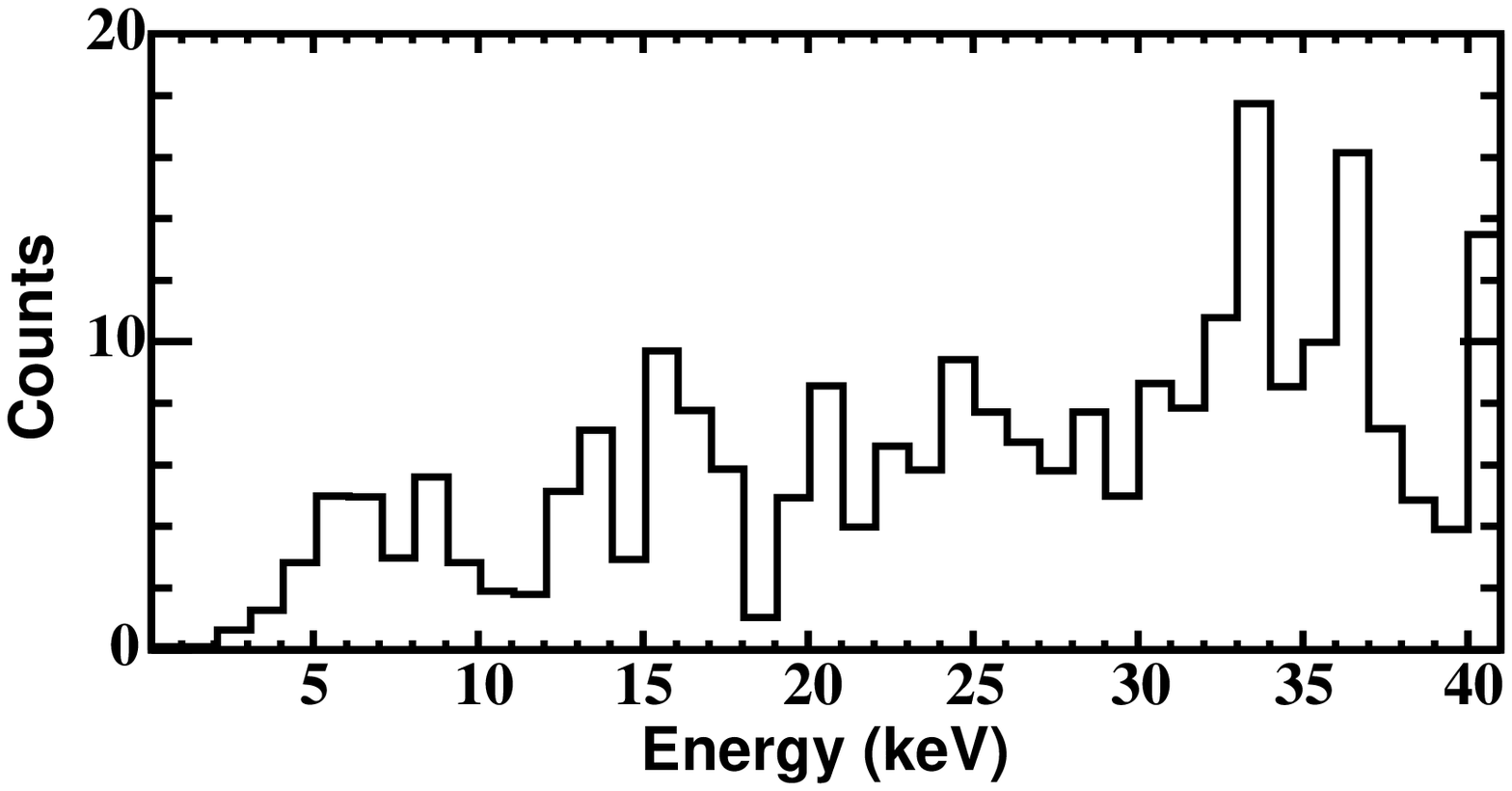}
\caption{\label{fig:specd5}Low energy spectra in the middle cell (solid line) and top cell (dotted line). 
The main expected lines coming from the low activity \Co source are indicated. 
The main structures at 7.3 and 13.6 keV are clearly seen as well as
two other structures at 20 and 27 keV in the middle cell spectrum. The background coming from
cosmic muons can be estimated from the top cell spectrum (dotted line).
The down-side plot shows the expected background coming from the 14.4, 122 and 136 keV \gams~and cosmic muons
with a Geant4~\cite{geant4} simulation reproducing the main features of the prototype
as explained in section~\ref{sec:prototype}. It can be directly compared to the spectrum measured on the top cell.
In the 1-40 keV range, 97\% are coming from cosmic muons, whereas only 
3\% are attributed to 14.4, 122 and 136 keV $\gamma$ interactions in \hetrois or in the copper walls.}
\end{center}
\end{figure}

Data at T$\sim$100~\muK~have been acquired during 15.5 hours with
a 5 cm thick lead shielding mounted around the nuclear demagnitization cryostat.
The data analysis procedure described above has been applied to the data. 
To obtain the low energy electron spectrum and its background,
we concentrate on the analysis on the top and middle cells since the data from the bottom one is much noisier.
Fig.~\ref{fig:rawdata} shows a typical sample of data from the cell containing the source,
with 100 ms sampling time.
The calibration coefficient has been estimated, as discussed below, to be ${\rm 2.4\times10^{-4}}$Hz/keV.
Raw data consist of a series of peaks corresponding to the energy released 
by the interaction of a particle inside the cell.
The rising ($\sim$1s) and relaxing ($\sim$5s) times are
the same for all the peaks. 
The data baseline, which is very stable, is related to very low frequency fluctuations on the temperature of
the \hetrois bath. Cosmic muons as well as electrons coming from the source are present in the data
in the proportion of 0.7 {\rm min$^{-1}$} and 3.8 {\rm min$^{-1}$} respectively.
The bottom plot shows a zoom on very low energy events, from 5 to 20 keV,
coming from the \Co source which are well above the background level. 
Fig.~\ref{fig:rawdata} shows 
the performance of the method on low energy events from 7 to 30 keV. 
The fit to the data (dashed line) as well as the residuals (dotted line) are presented.
We observe that a very good fit to the data is obtained with the mean of the residuals smaller than 1 keV. \\
From this analysis we have obtained the low energy electron
spectrum from the middle cell and the muon background spectrum from the
top cell. Both are presented on Fig.~\ref{fig:specd5} in solid and dotted lines respectively.
The spectrum of the middle cell presents a set of peaks, 
two of which correspond to the K 
and L shell internal conversion electrons of the 14.4 keV nuclear transition which are
respectively at 7.3 and 13.6 keV.
By matching the energy of these two peaks to their nominal value, we have obtained an absolute calibration factor
to convert from Hz to keV.
This calibration factor shows no evolution with energy. Indeed, cosmic muon distribution 
corresponding to energy peaks of about 67 keV is located at the expected value with such a factor.
The Auger electrons corresponding to the K and L shells contribute to energies of 0.6 and 5.5 keV respectively. The pile-up 
have been determined at 12.8 and 14.2 keV due to the superposition of 
conversion electrons with their corresponding
Auger electron whose fluorescence yields are ${\rm \omega_{K}\,=\,0.352}$ and ${\rm \bar{\omega}_L\,=\,0.006}$.
The photon emission is composed of 6.4 and 7.1 keV \xrays~and 14.4, 122 and 136 keV \gams.
As they interact mainly in the gold foil,
photoelectrons can escape. The L-shell electrons from \Au may pile-up with the K 
or L shell conversion electrons 
providing extra contributions at 21.7 and 27.9 keV.
Electrons emitted in a 4$\pi$ solid angle 
with energies lower than 40 keV, present a range smaller than 370 \microns and thus 
the electron contributions are to be detected in 
the cell containing the \Co source. These main expected electron lines in the 1 to 40 keV energy range,
coming from the low activity \Co source are listed in Table~\ref{table1} as well as their 
energy and the relative probability of the energy to be released in the cell.
Simulations based on the nuclear desintegration scheme of the source and on the geometry of the 
prototype permit to calculate such a probability.\\
\indent The electron spectrum in Fig.~\ref{fig:specd5} has been obtained 
with cut of 5 in the signal to noise ratio to keep best fitted peaks.
 The fact to see that the electron lines are clearly detected confirms
the very low background coming from the Compton interaction of the 122 and 136 keV \gams~
from the \Co source. 
Simulations including the electron, \gam~ and muon contributions have been produced for different source
activities. 
The analysis of these data, using the above processing procedure, 
allows us to give an estimate of the activity of the source which is 0.06 $\pm$ 0.01 Bq.
This is obtained by comparing the number of counts in the simulated and experimental electron spectra 
in the 1 to 18 keV energy range accounting for the efficiency of the analysis method.
The efficiency has been 
estimated as a function of the signal to noise ratio cut.
For signal to noise ratio higher than 5, the efficiency is
35\% in the 1-10 keV energy range and 86\% in 10-18 keV one. 
The experimental number of counts in both energy ranges are consistent within 5\%  with the
predictions obtained from Table~\ref{table1} including the efficiency of the analysis 
for an activity of 0.06 Bq.\\ 
\indent The background to the electron line detection has been measured in the top cell.
As the top cell cannot see the electrons, the spectrum obtained in it is composed of 
\gams~and muons. It can be qualitatively 
compared, both in shape and total number of counts, to the simulated one 
presented on the bottom plot of Fig.~\ref{fig:specd5}. 
The latter has been obtained from 
simulations of \gams~and cosmic muons as described in section~\ref{sec:prototype}.
This spectrum is dominated by cosmic muons crossing the cell on peripheral tracks
indicating that the experimental background is dominated by muons as well.
In addition, no 8 keV photoelectric line in the background spectrum is observed, indicating 
the transparency of \hetrois~to the X-rays coming from the copper walls.
In the 1 to 6 keV range, seven low energy background events have been detected. 
Cross correlating the signals of the top and middle cells, five of those are clearly identified as
cosmic muons. This confirms the efficiency of the use of cell correlation to reject background events.
Two events remain after rejection.
They are events which would mimick a \neut~event for 
this design and matrix size. 
They correspond probably to cosmic muons crossing only the top cell on peripheral tracks.
A better rejection could be achieved using a matrix with a larger number of cells
as it has been proposed in Ref.~\cite{nima}.
\section{\label{sec:concl}Conclusion}
This letter presents first results on the detection of 
low energy events with the MACHe3 three cell prototype.  
Electrons in the keV range coming from a \Co source embedded inside one
of the cells have been detected in superfluid \hetro-B 
cooled down to $\sim$~100 \muK. 
We succeed in 
separating the K and L shell conversion electrons coming from the 14.4 keV nuclear transition. 
We have also detected
Auger electrons as well as electron pile-up events. 
We have also shown that the cross correlation between cells can be efficiently used to reject 
background muons events.
From these results, we confirm that the background coming from the Compton interaction of the 122 and 136 keV \gams~
from the \Co source is very low in the energy range of interest.
Furthermore, the feasibility of the detection of low energy events of about 5.6 keV 
in superfluid \hetrois is demonstrated.
This is a decisive breakthrough for the project of non-baryonic dark matter search with
\hetrois based detectors.

\newpage

\begin{figure}[!ht]
\begin{center}
\scalebox{0.7}[0.6]{\input{prototype.pstex_t}}
\caption{\label{fig:prototype} A view of the multicell MACHe3 prototype. Each cell is a copper cylindric box of 500 
\micron~thick walls
filled with superfluid 
\hetrois at 100 \muK. A vibrating wire resonator (VWR) in NbTi forming a semi-loop of 4.5 \micron~
is inside each cell. The C cell
contains an extra 13 \micron~ VWR for calibration purpose. A very low activity \Co source ($\lesssim$ 0.1~Bq) has been embedded in the B
cell on a 25 \micron~thick gold foil spot-welded to the wall.  A 200 \micron~ diameter hole permits
 thermal relaxation of the quasiparticles.}
\end{center}
\end{figure}

\begin{center}
\begin{table}[hb]
\begin{center}
\begin{tabular}{ccc}
\hline
\hline
Electron&Energy&Relative\\
lines&(keV) &probability\\
\hline
\hline
e$_{A_L}$&0.6--1.8&16.2\%\\
e$_{A_K}$&5.5&13.8\%\\
e$_{C_K}$&7.3&35.9\%\\
e$_{C_K}$+e$_{A_K}$&12.8&17.4\%\\
e$_{C_L}$&13.6&13.14\%\\
e$_{C_L}$+e$_{A_L}$&14.2--15.4&0.01\%\\
e$_{C_M}$&14.4&0.01\%\\
e$_{C_K}$+e$_{A_L}(^{197}Au)$&21.7&0.002\%\\
e$_{C_L}$+e$_{A_L}(^{197}Au)$&27.9&0.001\%\\
\hline
\hline
\end{tabular}
\end{center}
\caption{\label{table1}Low energy electron lines, in the 1 to 40 keV energy range, coming from the \Co source with their corresponding
energy taken into account the possible pile-up. 
The third column presents the relative probability of the energy to be released in the cell. This probability
is obtained by simulating completely the nuclear desintegration scheme of the \Co source accounting for the geometry
of the prototype.}
\end{table}
\end{center}


\begin{thebibliography}{00}
\bibitem{wmap}D.N.~Spergel {\it et al.}, \APJS{148}{2003}{175}
\bibitem{archeops}A.~Beno\^{\i}t {\it et al.},\AA{399 No. 3}{2003}{L25}
\bibitem{tegmark}M.~Tegmark {\it et al.}, \PRD{69}{2004}{103501}
\bibitem{snap}S.~Perlmutter {\it et al.}, \APJ{517}{1999}{565}
\bibitem{jungman}G.~Jungman {\it et al.}, \PR{267}{1996}{195}
\bibitem{edelweiss}S.~Marnieros {\it et al.}, \NIMA{520}{2004}{101}
\bibitem{cdms}D.S.~Akerib {\it et al.}, \PRL{93}{2004}{211301-1}
\bibitem{lancaster}G.~Pickett, Proc. Second european 
worshop on neutrinos and dark matters detectors, 
Ed. L. Gonzales-Mestres and D. Perret-Gallix, 
Editions Fronti\`eres, 1988, p. 377.
\bibitem{bunkov}Yu.~Bunkov {\it et al.}, Proc. International 
Workshop Superconductivity and Particles Detection, 
Ed.  T. Girard, A. Morales and G. Waysand, 
World Scientific, 1995, p. 21.
\bibitem{dm2000}D.~Santos {\it et al.}, Proc. of the
Fourth International Symposium on Sources 
and Detection of Dark Matter and Dark Energy in the Universe, February 2000, 
Marina Del Rey (CA, USA), Ed. D.B. Cline, Spinger, astro-ph/005332
\bibitem{nima}F.~Mayet {\it et al.}, \NIMA{455}{2000}{554}
\bibitem{4micc}E.~Moulin {\it et al.}, Proc. of the
Fourth International Conference on Where Cosmology and Fundamental 
Physics Meet, Marseille, June 2003, astro-ph/0309325
\bibitem{plbf}F.~Mayet {\it et al.}, \PLB{538}{2002}{257}
\bibitem{nature95}C.~B\"auerle {\it et al.}, \NATURE{382}{1995}{332}
\bibitem{prb98}C.~B\"auerle {\it et al.},\PRB{57}{1998}{14381}
\bibitem{geant4}S.~Agostinelli {\it et al.}, \NIMA{506}{2003}{250}
\bibitem{grieder}P.~Grieder,Cosmic Rays at Earth: Researcher's Reference Manual and Data Book,
 Elsevier Science, 2001
\bibitem{collin}E.~Collin {\it et al.}, {\it submitted to Astropart. Phys.} 
\end{thebibliography}
\end{document}